\documentclass[12pt, letterpaper]{article}

\usepackage{graphicx}
\usepackage{amsmath}
\usepackage{mathrsfs}
\usepackage{fullpage}

\newcommand{\WW}{\mathscr W}
\newcommand{\Dray}{\cite{Dray:1985yt}}
\newcommand{\us}[1]{#1}
\newcommand{\nline}{$45^\circ$}

\begin{document}

\title{The Effect of Negative-Energy Shells on the Schwarzschild Black Hole}
\author{Jeffrey S Hazboun\\
Department of Physics\\
Oregon State University\\
Corvallis, OR  97331\\
\texttt{jeffrey.hazboun@gmail.com}
\and
Tevian Dray\\
Department of Mathematics\\
Oregon State University\\
Corvallis, OR  97331\\
\texttt{tevian@math.oregonstate.edu}
}
\date{3 July 2009}

\maketitle

\begin{abstract}
We construct Penrose diagrams for Schwarzschild spacetimes joined by massless
shells of matter, in the process correcting minor flaws in the similar
diagrams drawn by Dray and 't~Hooft~\Dray, and confirming their result that
such shells generate a horizon shift.  We then consider shells with negative
energy density, showing that the horizon shift in this case allows for travel
between the heretofore causally separated exterior regions of the
Schwarzschild geometry.  These drawing techniques are then used to investigate
the properties of successive shells, joining multiple Schwarzschild regions.
Again, the presence of negative-energy shells leads to a causal connection
between the exterior regions, even in (some) cases with two successive shells
of equal but opposite total energy.
\end{abstract}

\section{Introduction}

Penrose diagrams are commonly used to describe the global structure of
spacetimes.  The mathematically rigorous construction of these diagrams, first
given by Penrose~\cite{Penrose:1965}, involves a careful look at the
asymptotic behavior of spacetime, as is well summarized
in~\cite{Esposito:1977w}.
Such conformal diagrams are well adapted for showing the global structure of
spacetimes containing null shells, and many examples of such Penrose diagrams
can be seen in~\cite{Barrabes:2003nh}.  Dray and 't~Hooft presented similar
diagrams in~\Dray, which described how spherically symmetric shells of
massless matter affect Schwarzschild spacetime.

Although Penrose diagrams have been proposed for more involved spacetime
structures, such as the collapse of stars and the evaporation of black
holes~\cite{Carroll:2003sg}, these are often presented without rigorous
mathematical construction.  Here we will extend the work of Dray and 't~Hooft
by carefully drawing Penrose diagrams in one global set of coordinates.  This
will allow us a visual means of investigating the structure of these
Schwarzschild spacetimes joined by shells, as is particularly well
demonstrated by our investigations of shells with negative energy density and
of successive shells.

The paper is organized as follows.  In Section~\ref{DtH} we present the
Penrose diagrams given in~\Dray, which we then redraw more accurately in
Section~\ref{Shifted}.  In Section~\ref{Negative}, we consider the effect of
shells with negative energy density, showing that such shells can be used to
construct traversable wormholes.  In Section~\ref{Successive}, we consider the
effect of successive, non-intersecting shells, paying particular attention to
the case of concentric shells of equal but opposite total energy.  Finally, in
Section~\ref{Discussion}, we discuss our results.

\section{Joined Schwarzschild Penrose Diagrams}
\label{DtH}

\begin{figure}[t]
\begin{center}
\includegraphics[width=3in]{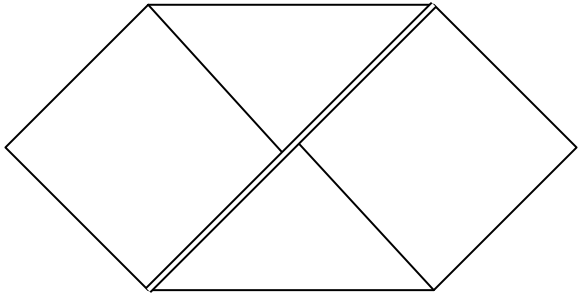}
\caption{\textbf{The Penrose diagram of a particle crossing the horizon as
drawn in~\cite{Dray:1984ha}} Here we see that the Dray-'t~Hooft solution has a
shift in the horizon as one crosses the world line of the shell.  Note that
requiring the outer boundary to be the same as for ordinary Schwarzschild, as
drawn here, prevents the shifted horizon from being drawn at~\nline.}
\label{ShiftedHorizon}
\end{center}
\end{figure}

Dray and 't~Hooft~\Dray\ constructed solutions of the Einstein field equation
which represent two separate Schwarzschild spacetimes joined along spherically
symmetric shells of massless matter.  These shells are described by
stress-energy tensors of the form
\begin{align}
T_{uu} & =\frac{\kappa}{4\pi}\delta(u) & m_1 & =m_2 \\
T_{uu} & =\frac{m_1(m_2-m_1)}{\pi \alpha r^2}\delta(u-\alpha) & m_1 & \neq m_2
\label{StressEnergy}
\end{align}
where $u$ is a null Kruskal-Szekeres coordinate, $u=0$ corresponds to the
horizon, and the shell is located at $u=\alpha$.  One of the more interesting
features of spacetimes joined in such a manner, first shown explicitly
in~\cite{Dray:1984ha} and drawn in Figure~\ref{ShiftedHorizon}, is that there
is a shift in the horizon as an observer crosses over the shell.  This result
is easily explained in the case where the shell separates two Schwarzschild
regions with different masses, since $r=2m_1$ and $r=2m_2$ will meet the
shell, on their respective sides, at different points.  Surprisingly this
shift also occurs in the case where both regions are of equal mass and the
shell is at the horizon.

\begin{figure}[t]
\begin{center}
\includegraphics[width=6in]{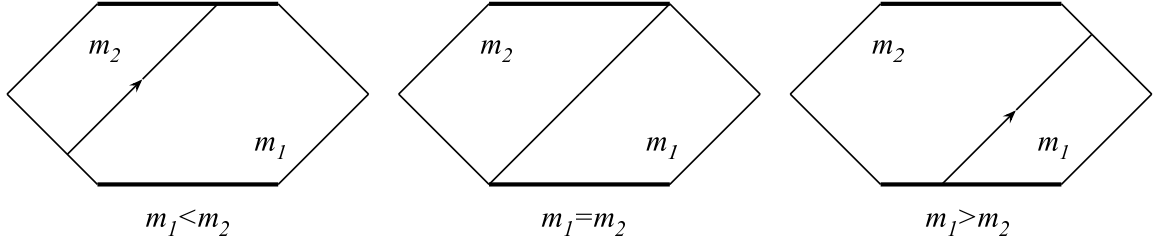}
\caption{\textbf{Dray-'t~Hooft Conformal Diagrams:} These diagrams depict the
three cases considered by Dray and 't~Hooft for a single null shell~\Dray.
The first diagram represents a shell of positive energy density symmetrically
collapsing to the singularity.  The center picture is a shell at the horizon,
and the last diagram shows a shell expanding from the singularity out to
infinity.}
\label{DraytHooft3}
\end{center}
\end{figure}

A careful look at the Penrose diagram in Figure~\ref{ShiftedHorizon} shows
that it is impossible to draw both the shifted and unshifted horizons as null
lines while preserving the boundary.  The horizons are null lines, and should
therefore be kept at \nline\ as we cross the boundary.  The center picture in
Figure~\ref{DraytHooft3} is therefore drawn incorrectly.  The off-horizon,
unequal-mass Penrose diagrams are correct, if we assume each region is drawn
in its own set of coordinates, however, they are misleading as drawn.  We will
show below that drawing the entire diagram with one set of coordinates changes
the shape of the singularity, which is the only line not drawn along a null
trajectory.  By drawing the pictures, and especially the boundary, more
accurately, it is possible to obtain physical information directly from the
diagrams, such as the relative shift in horizons and the relative positions of
horizons.

\section{Shifted Diagrams}
\label{Shifted}

We begin by reviewing the construction of the spacetimes in~\Dray.  In order
to glean as much physical understanding from these Penrose diagrams as
possible, we shall be pedantic about the use of one coordinate system to draw
the entire spacetime.  Throughout this paper, $u_1$ and $v_1$ will refer to
the coordinates of the region where $u<\alpha$ and $u_2$ and $v_2$ will refer
to the coordinates of the region where $u>\alpha$.  We first assume that the
coordinates in Region~2 can be expressed as functions of $u_\us{1}$ and
$v_\us{1}$.  As shown in~\Dray, by requiring the metric to be continuous at
the join we get two relations,
\begin{equation}
\frac{dv_2}{dv_1}
  = \frac{m_1^3}{u'_2(\alpha)m_2^3} e^{-\frac{r}{2m_1}
	+ \frac{r}{2m_2} \mid_{u_1=\alpha}}
\label{vde}
\end{equation}
\begin{equation}
\frac{\alpha}{m_1} = \frac{u_2(\alpha)} {m_2 u_2'(\alpha)}
\label{u2u1}
\end{equation}
which can be used to find $u_2$ and $v_2$ as functions of $u_1$ and $v_1$,
respectively.  It is important to note that~\eqref{vde} in principle yields a
global relationship between $v_\us{1}$ and $v_\us{2}$, whereas~\eqref{u2u1}
refers to a boundary condition that applies only at the shell.

\subsection{Equal Mass Diagrams}

In the case of a shell at the horizon, \eqref{vde} and~\eqref{u2u1} can be
simplified extensively, resulting in the expressions
\begin{eqnarray}
u_\us{2} &=& u_\us{1} \nonumber\\
v_\us{2} &=& v_\us{1}+\kappa
\label{EqualMass}
\end{eqnarray}
where $\kappa$ is the constant from the $T_{uu}$ component of the stress
energy tensor in~\eqref{StressEnergy}.  We have also assumed the relationship
in~\eqref{u2u1} is a global relationship, and linear in $u_\us{1}$.  
Penrose diagrams are traditionally drawn using the canonical asymptotic
transformation
\begin{equation}
U=\arctan(u),\quad V=\arctan(v)
\label{Conformal}
\end{equation}
By making the substitutions from \eqref{EqualMass} before doing the asymptotic
coordinate transformation on the spacetime, so that
\begin{equation}
U=\arctan(u_\us{2}),\quad V=\arctan(v_\us{1}+\kappa)
\label{ConformalShifted}
\end{equation}
we obtain a Penrose diagram that differs from the standard ``straight''
Schwarzschild conformal diagram.

\begin{figure}[t]
\begin{center}
\begin{minipage}{0.47\linewidth}
\includegraphics[width=3in]{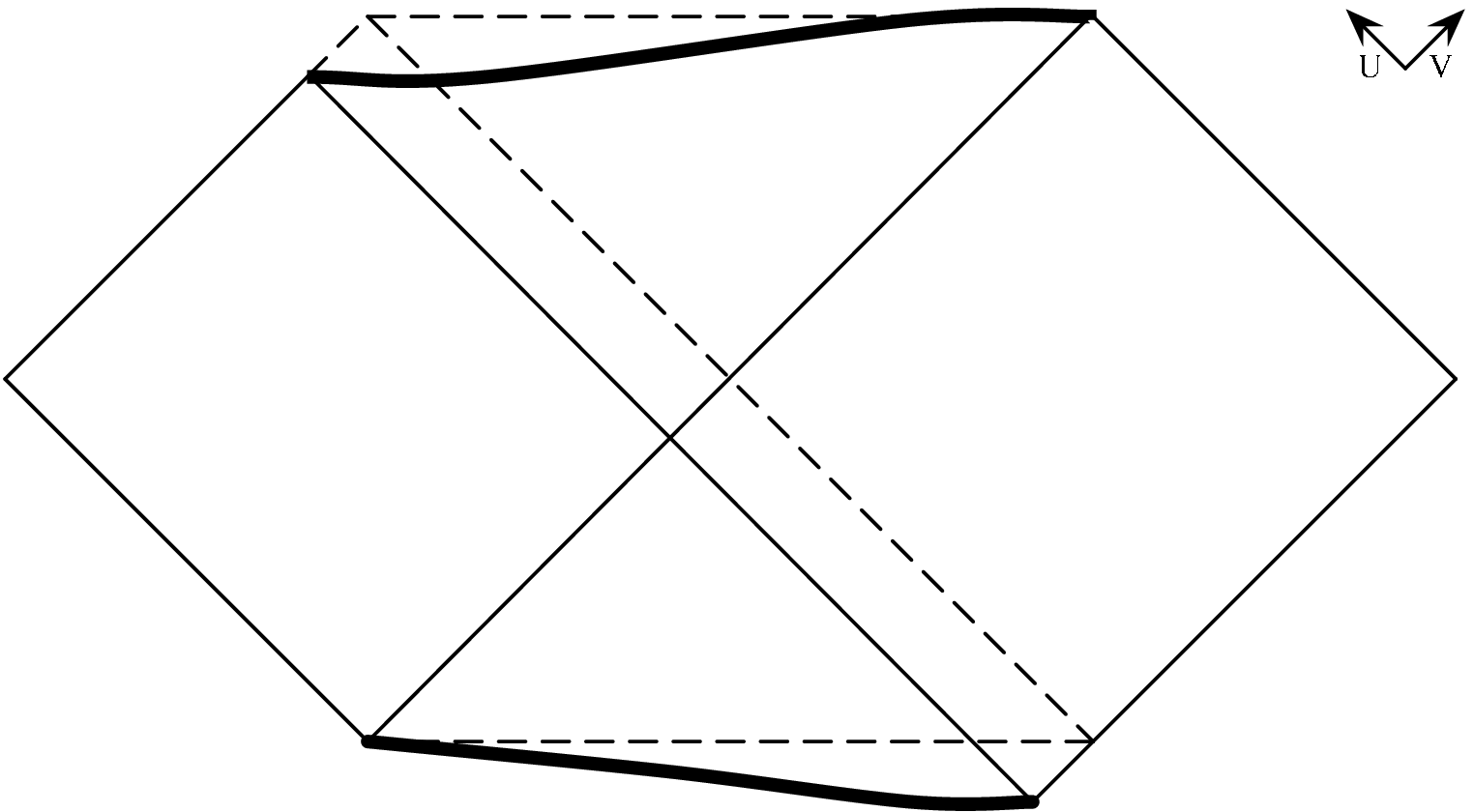}
\end{minipage}
\begin{minipage}{0.47\linewidth}
\includegraphics[width=3in]{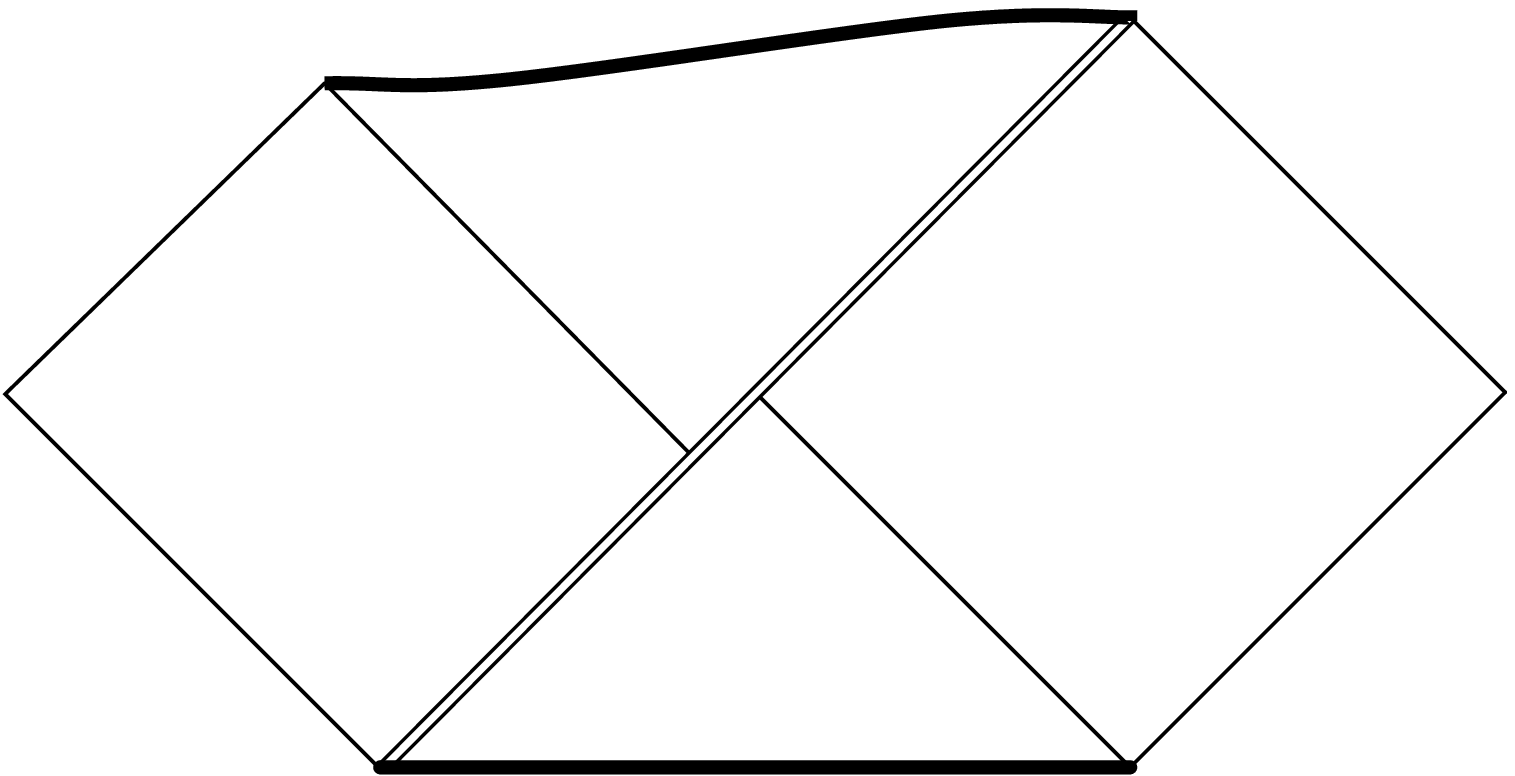}
\end{minipage}
\caption{\textbf{Equal Mass Diagrams:} The first diagram in this figure is the
Penrose diagram drawn for a spacetime with coordinates $u_\us{2}$ and
$v_\us{2}$ as in equations~\eqref{EqualMass}.  The bold lines are the
singularities drawn in the shifted coordinates.  The second figure represents
the Dray-'t~Hooft spacetime with the shell represented by the double line at
the horizon.}
\label{Shift}
\end{center}
\end{figure}

The first diagram in Figure~\ref{Shift} shows the shifted coordinate system
drawn on top of the straight Penrose diagram for Schwarzschild spacetime; both
superimposed diagrams represent a Schwarzschild spacetime of the same mass.
The most striking change in the diagram is in the shape of the singularities,
which are no longer perfectly horizontal.  In the second diagram in
Figure~\ref{Shift}, we have split the spacetime along the location of the
shell, at the horizon, and joined it to a region with unshifted coordinates.
This accurately depicts the entire Dray-'t~Hooft spacetime, drawn in the
coordinates of the unshifted region.

The set of coordinates the diagrams are drawn in is of course arbitrary; we
could just as easily have drawn these pictures using the $u_\us{2}$ and
$v_\us{2}$ coordinates.  This would have the effect of curving the white hole
singularity instead of the black hole singularity, but the relative shift in
the horizon would be the same.

\subsection{Unequal Mass Diagrams}

\begin{figure}[t]
\begin{center}
\begin{minipage}{0.47\linewidth}
\includegraphics[width=3in]{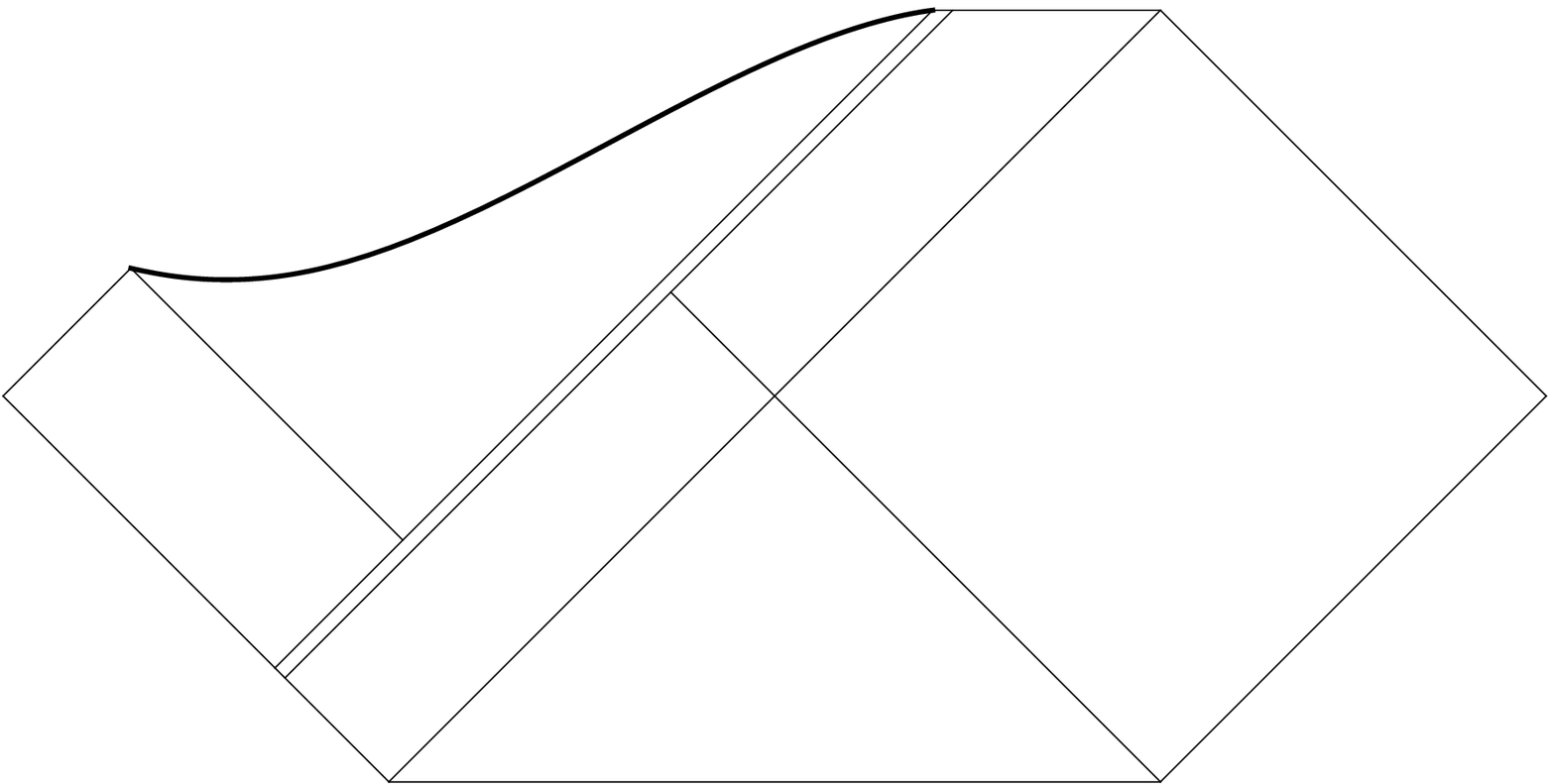}
\end{minipage}
\begin{minipage}{0.47\linewidth}
\includegraphics[width=3in]{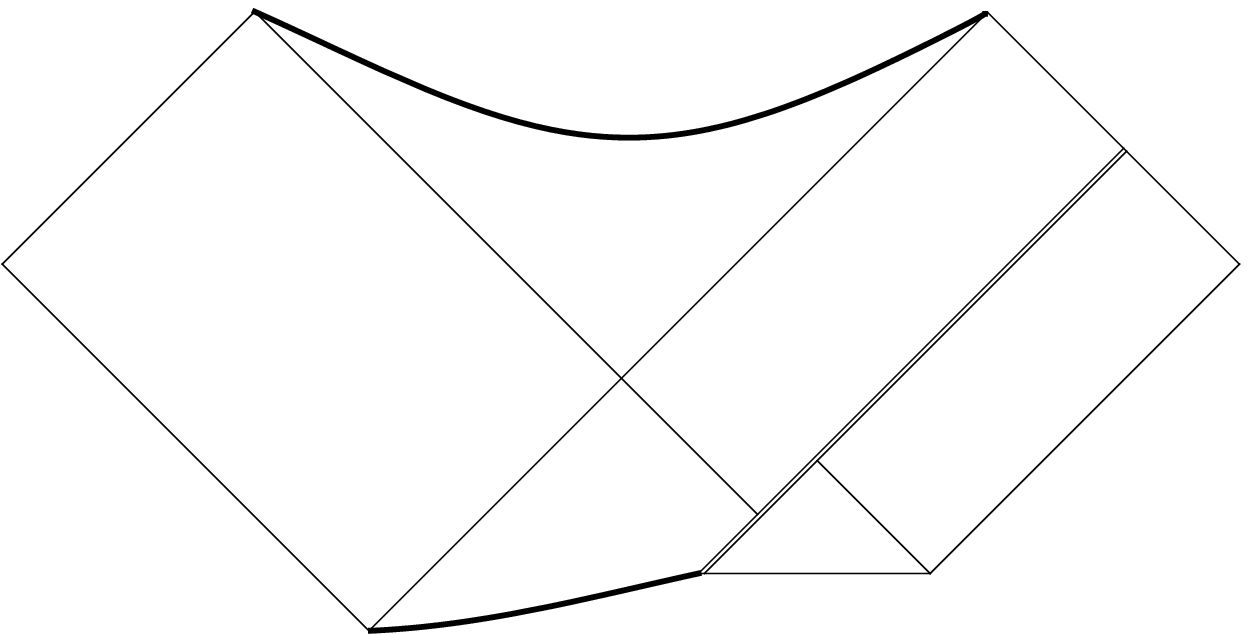}
\end{minipage}
\caption{\textbf{Unequal Mass Diagrams:} On the left is the diagram for a
spacetime with a shell at $\alpha=\frac{1}{2}$, and
$m_\us{2}=\frac{5}{4}m_\us{1}$.  On the right is a diagram for the case where
$\alpha=-\frac{3}{2}$, and $m_\us{2}=\frac{4}{5}m_\us{1}$}
\label{UnequalMass}
\end{center}
\end{figure}

The case of a shell that is not at the horizon is more complicated than the
equal mass (at the horizon) case.  Going back to the general
relations~\eqref{vde} and~\eqref{u2u1} and solving for the Region~2
coordinates as functions of the Region~1 coordinates, we get
\begin{equation}
u_2 = u_1 + \alpha \left(\frac{m_\us{2}}{m_\us{1}}-1\right)
\label{UnequalU}
\end{equation}
\begin{equation}
v_\us{2}(v_\us{1})
  = \left(m_\us{1}\: \boldsymbol{\WW}\negthickspace\!
	\left(-\frac{v_\us{1}\alpha}{e}\right)+2(m_\us{1}-m_\us{2})\right)
	\frac{m_\us{1}}{\alpha m_\us{2}^2 u_\us{2}'(\alpha)}
	e^{\frac{m_\us{1}}{m_\us{2}}
	\left(\boldsymbol{\WW}\!\left(-\frac{v_\us{1}\alpha}{e}\right)\right)}
\label{UnequalV}
\end{equation}
where $\WW$ is the Lambert W function (also known as the Product Log).  The
integration constant, found by setting $v_2(v_1)=v_1$ at the singularity, is
identically zero.  There is a freedom in the choice of $u_2(u_1)$, and we have
chosen a simple linear relationship that satisfies~\eqref{u2u1}.
Also,~\eqref{UnequalU} again implicitly assumes that the relation
in~\eqref{u2u1} is true for all values of $u_1$.  These functions are then
used to draw the Penrose diagram in Region~2.  The singularity in Region~2,
where $u_2v_2=1$, is computed as the nested function $u_1(u_2(v_2(v_1)))$.
The shifted horizon is drawn as a null line starting at $r=2m_2$ on the shell.
Figure~\ref{UnequalMass} shows the shifted diagrams for shells with positive
energy density that are collapsing or expanding, respectively.

\begin{figure}[t]
\begin{center}
\includegraphics[width=3in]{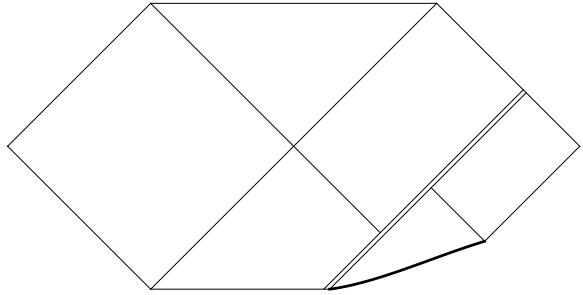}
\caption{\textbf{Negative $\alpha$ diagram drawn in $u_\us{2}$, $v_\us{2}$
coordinates:} Here we have drawn the entire diagram in the Region~2
coordinates.  By using these coordinates, there is no need to use an arbitrary
function.}
\label{NegAlpha}
\end{center}
\end{figure}

The second diagram in Figure~\ref{UnequalMass}, with $\alpha<0$, results in a
shifted white hole singularity.  This singularity, however, does not lie
within the light cone of the shell, and cannot therefore be described by the
metric continuity equations,~\eqref{vde} and~\eqref{u2u1}, that we have
previously used to construct the shifted coordinates.  In
Figure~\ref{UnequalMass}, we have chosen a linear relationship for
$v_\us{2}(v_\us{1})$.  In fact, the Lambert W function becomes complex at
$\frac{1}{\alpha}$, mathematically dictating that the original coordinate
relationship~\eqref{UnequalV} is no longer useful.  The choice of coordinate
extension used is completely arbitrary, so long as modest continuity
conditions hold.  We can nonetheless use the relationships in~\eqref{UnequalU}
to fully describe this case if we instead treat the $u_\us{1}$ and $v_\us{1}$
coordinates as functions of $u_\us{2}$ and $v_\us{2}$.  Figure~\ref{NegAlpha}
shows a Penrose diagram for a shell with $\alpha<0$, with the Region~1
singularity drawn in $u_\us{2}$ and $v_\us{2}$ coordinates.  Similar comments
apply to the case $\alpha>0$, with the roles of the regions reversed.

These diagrams reveal some important characteristics of these spacetimes.  As
shown in Figure~\ref{UnequalMass}, when a shell passes over an observer, their
relative position with respect to the horizon transverse to the shell is
changed.  In the positive $\alpha$ case (the first diagram in
Figure~\ref{UnequalMass}), an observer in the (left) asymptotic region may
even be shifted into the black hole.

\begin{figure}[t]
\begin{center}
\begin{minipage}{0.4\linewidth}
\begin{center}
\includegraphics[width=2in]{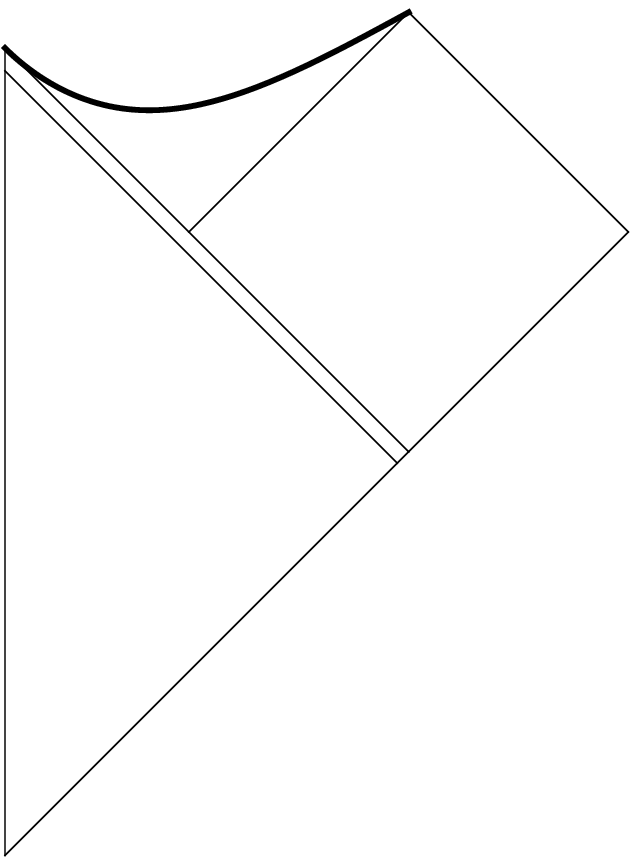}
\end{center}
\end{minipage}
\begin{minipage}{0.4\linewidth}
\begin{center}
\includegraphics[width=2in]{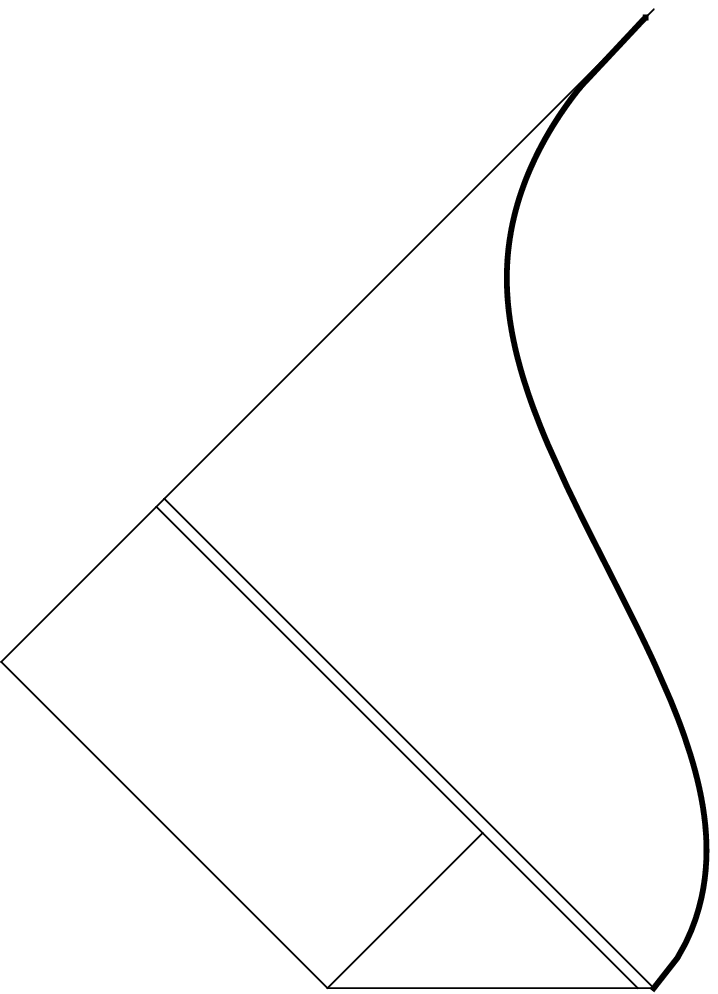}
\end{center}
\end{minipage}
\caption{\textbf{Minkowski$\leftrightarrow$Schwarzschild Mass Diagrams:}
On the left is the diagram for a spacetime with a shell at
$\alpha=\frac{1}{2}$.  On the right is a diagram for the case where
$\alpha=-1$.}
\label{Flat}
\end{center}
\end{figure}

For completeness, we apply similar techniques to the next case considered
in~\Dray\ and look at the Penrose diagrams when $m_\us{1}=0$ or $m_\us{2}=0$.
Here we use a shell with a trajectory along the null Kruskal-Szekeres
$u$-direction, in order to retain the standard orientation of the Penrose
diagram for Minkowski space.  Again we see that the singularities are no
longer straight, as shown in Figure~\ref{Flat}.

In constructing Figure~\ref{Flat}, we have again used a linear function to
describe the relationship between the coordinates transverse to the shell,
which are now $v_1$ and $v_2$.  In the second diagram, the simplest linear
relationship that satisfies \eqref{u2u1} is
\begin{equation}
v_2=\frac{2m_1 v_1}{\alpha}
\label{FlatV2V1}
\end{equation}
Note that the second diagram in Figure~\ref{Flat} is not changed when the mass
of Region 1 is changed.

\section{Negative Energy Shells and Wormholes}
\label{Negative}

Until this point, we have been careful to consider only spacetimes containing
shells with positive energy density.  Using our techniques, however, it is
elementary to study the effects of shells with negative energy density.
Looking explicitly at the energy density component of the stress-energy
tensor, transformed to the canonical Schwarzschild coordinates,
\begin{equation}
-T_{t}^{\;t}
  = \frac{(m_\us{2}-m_\us{1})}{4\pi r^2}\frac{|\alpha|}{\alpha}\delta(r-r_0)
\label{Ttt}
\end{equation}
we see that the sign of the energy density depends on both the sign of the
change in mass between the two regions and on the sign of $\alpha$.  This
differs from the similar expression given in~\cite{1989GReGr..21..741D}, where
the dependence on $\alpha$ has been omitted, since negative energy density
shells were not being considered.  The dependence on $\alpha$ arises because
the sign of $\alpha$ dictates whether the shell is expanding or contracting.
Note that we have not concerned ourselves with the ``how'' of negative energy
density, but only its effects.  For detailed discussions of negative energy in
general relativity, see Ford and Roman~\cite{Ford:1993bw}.

\begin{figure}[t]
\begin{center}
\begin{minipage}{0.47\linewidth}
\includegraphics[width=3in]{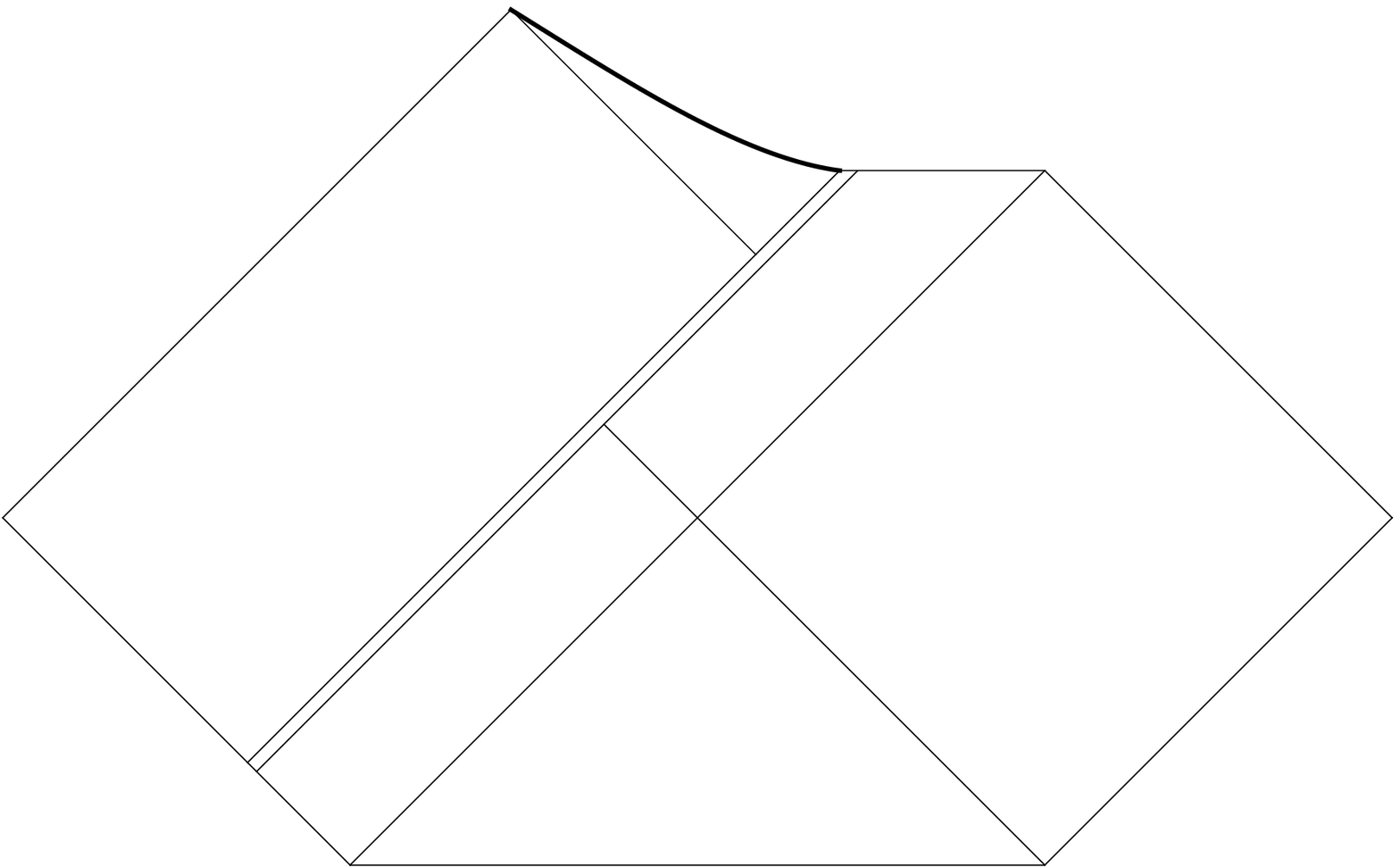}
\end{minipage}
\begin{minipage}{0.47\linewidth}
\includegraphics[width=3in]{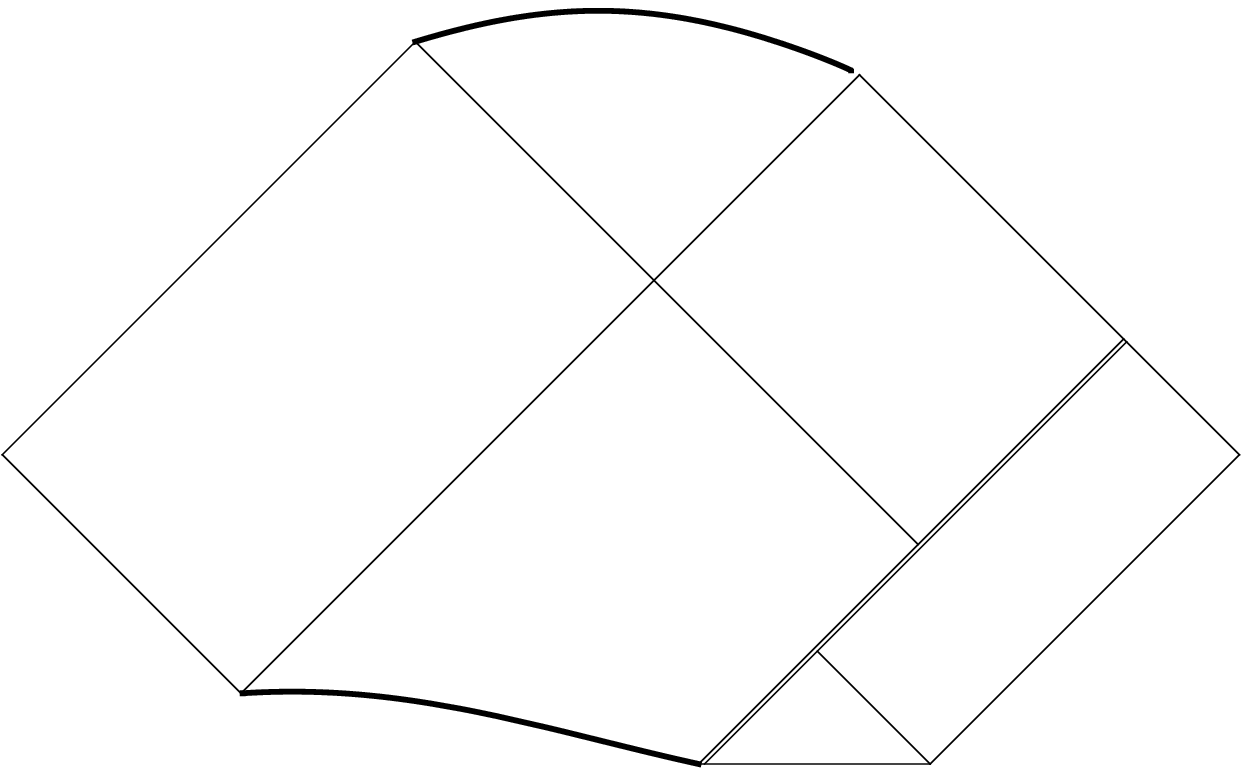}
\end{minipage}
\caption{\textbf{Unequal Mass Diagrams with Negative Energy Density Shells:}
The first diagram is for a spacetime with a shell at $\alpha=\frac{1}{2}$, and
$m_\us{2}=\frac{4}{5}m_\us{1}$.  The second diagram depicts the case where
$\alpha=-\frac{3}{2}$, and $m_\us{2}=\frac{5}{4}m_\us{1}$.  Note the shift has
opened a gap between the two $u_\us{1}$ and $u_\us{2}$ horizons, that a
time-like or light-like trajectory can pass through.}
\label{NegUnequal}
\end{center}
\end{figure}

When we draw diagrams with negative energy density shells, there are some
important changes in the structure of the spacetime that are easily seen by
looking at our Penrose diagrams.  In Figure~\ref{NegUnequal}, we see that the
horizons, in both cases, have shifted in the positive $v$ direction.  This
shift in the horizon allows a null geodesic to pass through the Einstein-Rosen
bridge to the other asymptotically flat region of the spacetime.  These
figures show this important attribute of the spacetime with little
computational effort.

Wormholes in spherically symmetric spacetimes have been discussed at length by
Morris and Thorne~\cite{Morris:1988cz,Morris:1988tu} and
Visser~\cite{Visser:1989kh,Visser:1989kg}.  Each of these authors
demonstrates that the resulting spacetime must have a stress-energy tensor
that compromises either the weak energy condition or the averaged weak energy
condition.  It is therefore not surprising that our simple wormholes require
negative energy density for their construction.

\section{Successive Shells}
\label{Successive}

Our techniques can be easily generalized to draw Penrose diagrams of
spacetimes with multiple Schwarzschild regions joined along successive shells.
Through most of this paper, and in particular in Figures~\ref{UnequalMass}
and~\ref{NegUnequal}, we have used the $u_\us{1}$, $v_\us{1}$ coordinates to
draw the joined Penrose diagrams.  In Figure~\ref{NegAlpha} we drew Penrose
diagrams using the $u_\us{2}$, $v_\us{2}$ coordinates.  It is straightforward
to combine two such diagrams, resulting in a spacetime that contains two
successive shells, drawn using the coordinates from the middle region.  These
spacetimes satisfy the boundary conditions in~\Dray\ at each shell, but the
diagrams can be obtained by combining the appropriate ``jigsaw puzzle pieces''
from Figures~\ref{UnequalMass} and~\ref{NegAlpha}.

\begin{figure}[t]
\begin{center}
\begin{minipage}{0.47\linewidth}
\includegraphics[width=2.75in]{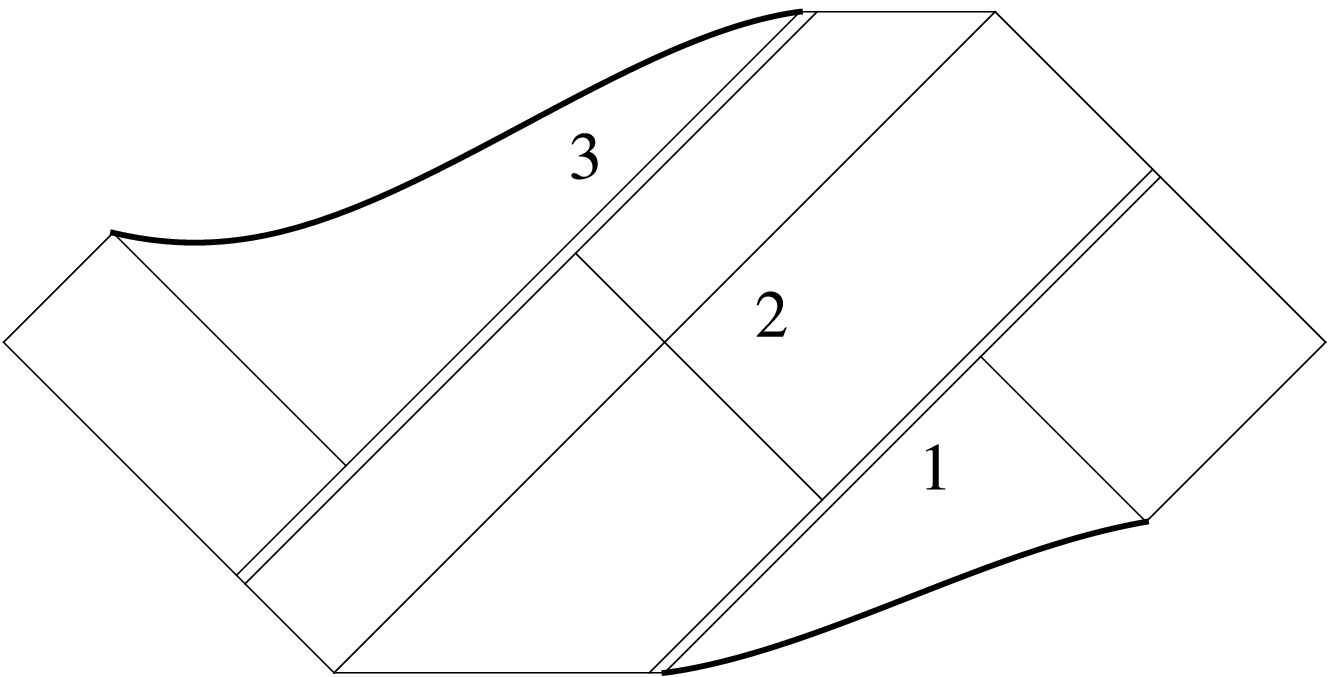}
\end{minipage}
\begin{minipage}{0.47\linewidth}
\includegraphics[width=2.75in]{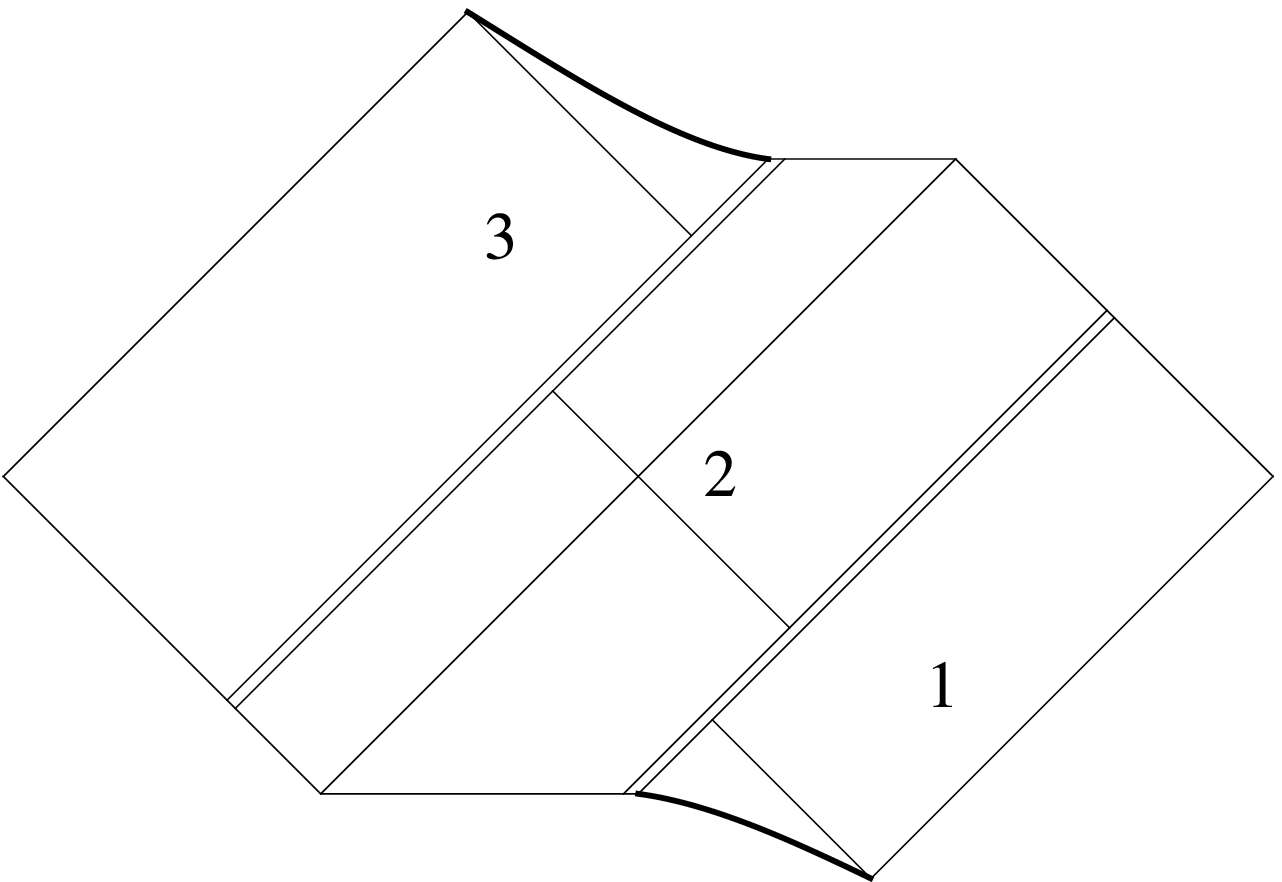}
\end{minipage}
\caption{\textbf{Two Successive Shells, Drawn in $u_\us{2}$ and $v_\us{2}$
coordinates:} Here the first diagram has two shells of positive energy
density, while the second diagram has two shells of negative energy density.}
\label{2Shellsu2v2}
\end{center}
\end{figure}

For example, in Figure~\ref{2Shellsu2v2} we have drawn two diagrams, each with
two shells.  The first diagram shows a spacetime with three Schwarzschild
regions, joined by two shells with positive energy density, while the second
diagram represents three Schwarzschild regions joined by two shells of
negative energy density.
	
\begin{figure}[t]
\begin{center}
\begin{minipage}{0.47\linewidth}
\includegraphics[width=2.75in]{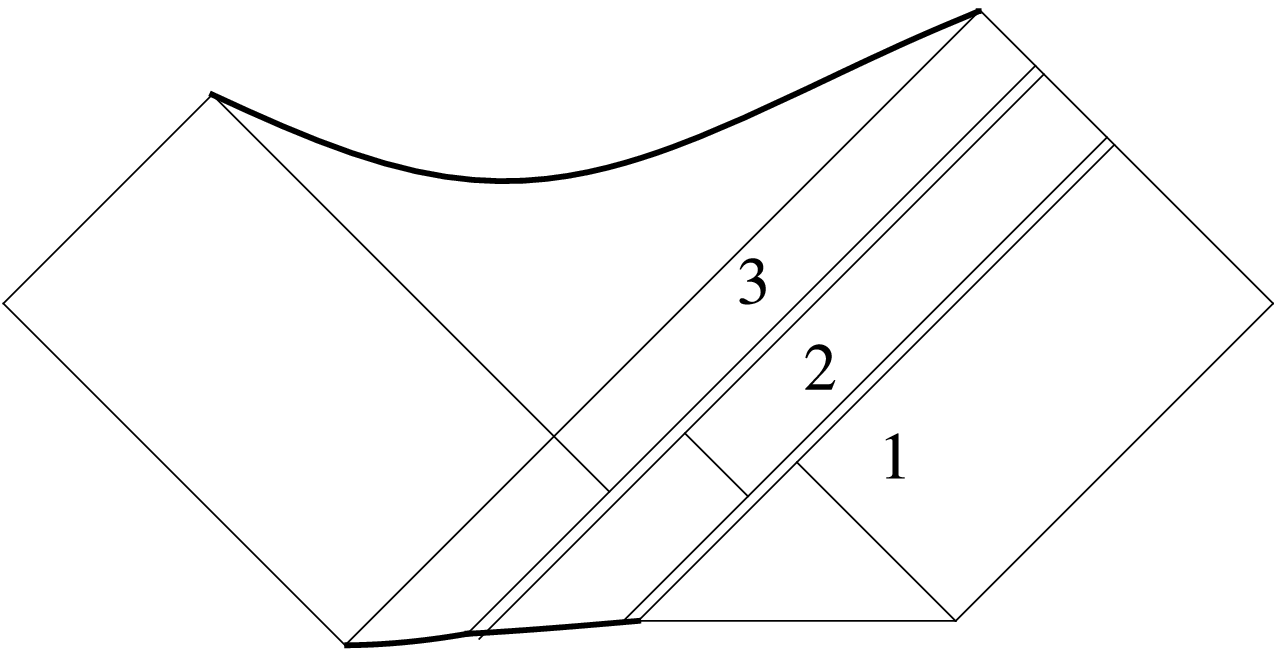}
\end{minipage}
\begin{minipage}{0.47\linewidth}
\includegraphics[width=2.75in]{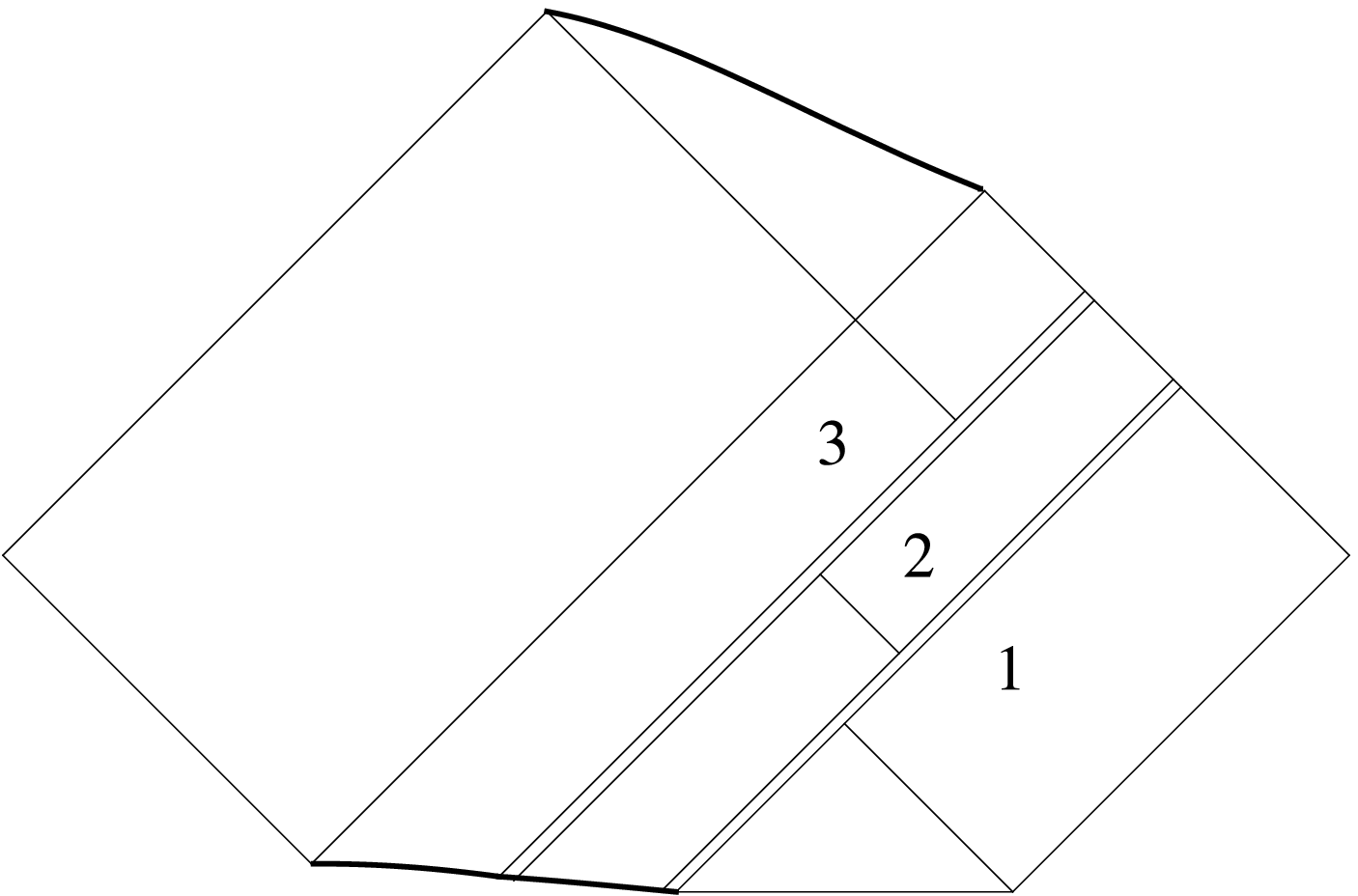}
\end{minipage}
\caption{\textbf{Two Successive Shells, Drawn in $u_\us{1}$ and $v_\us{1}$
coordinates:} The first diagram has two shells of positive energy density,
while the second has two shells of negative energy density}
\label{2Shellsu1v1}
\end{center}
\end{figure}

The same Penrose diagrams can also be drawn using the Region~1 coordinates.
In this case Region~3 is drawn by first relating the coordinates in Region~3
to those in Region~2 using~\eqref{vde} and~\eqref{u2u1}, and then inserting
the known functional dependence of the Region~2 coordinates on the Region~1
coordinates, obtaining expressions of the form
$u_\us{3}=u_\us{3}(u_\us{2}(u_\us{1}))$ and
$v_\us{3}=v_\us{3}(v_\us{2}(v_\us{1}))$.  While difficult to evaluate by hand,
\textsl{Mathematica} has no difficulty working with these composite functions,
leading to diagrams such as those shown in Figure~\ref{2Shellsu1v1}, drawn
entirely using the Region~1 coordinates.

\begin{figure}[t]
\begin{center}
\begin{minipage}{0.47\linewidth}
\includegraphics[width=3in]{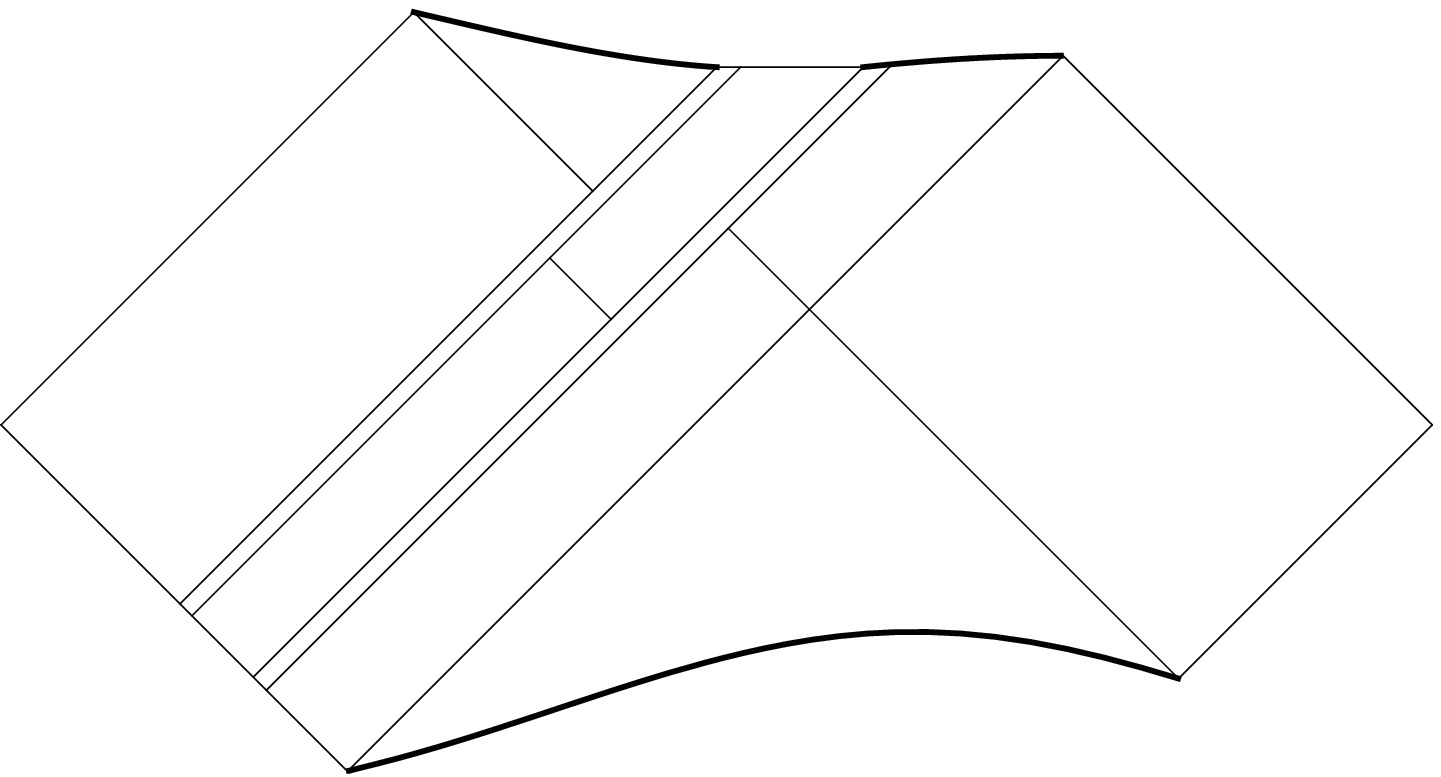}
\end{minipage}
\begin{minipage}{0.47\linewidth}
\includegraphics[width=3in]{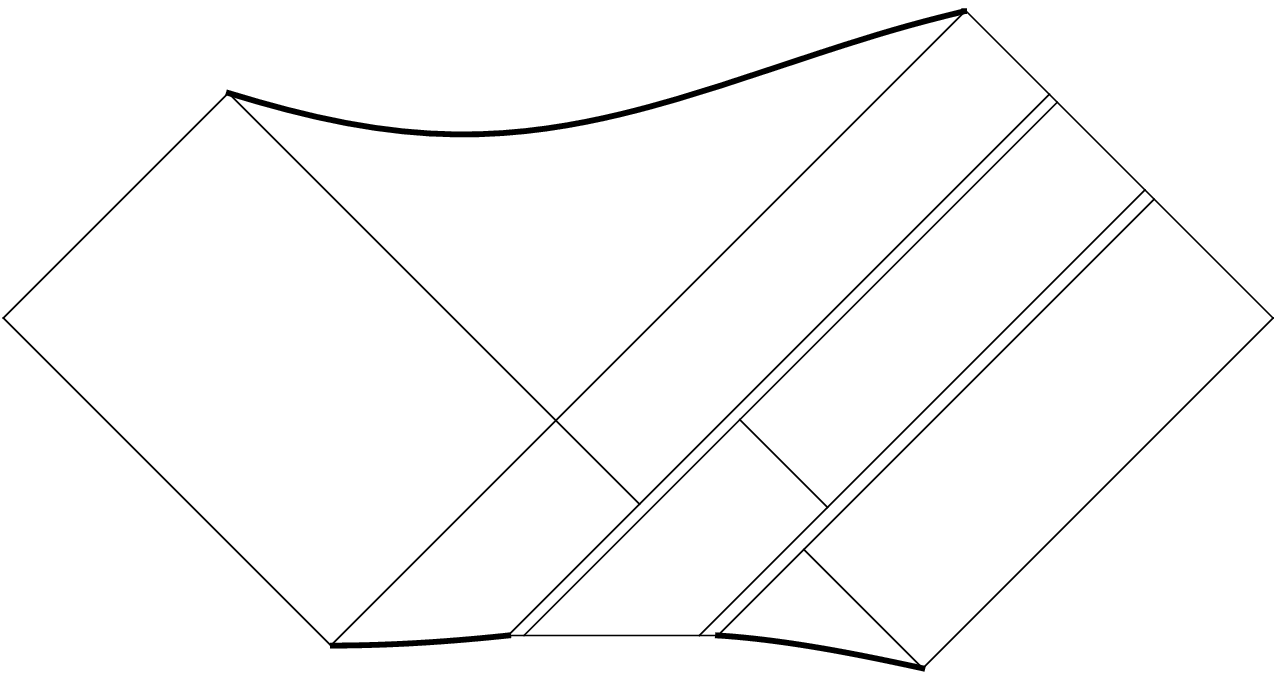}
\end{minipage}
\\[10pt]
\begin{minipage}{0.47\linewidth}
\includegraphics[width=3in]{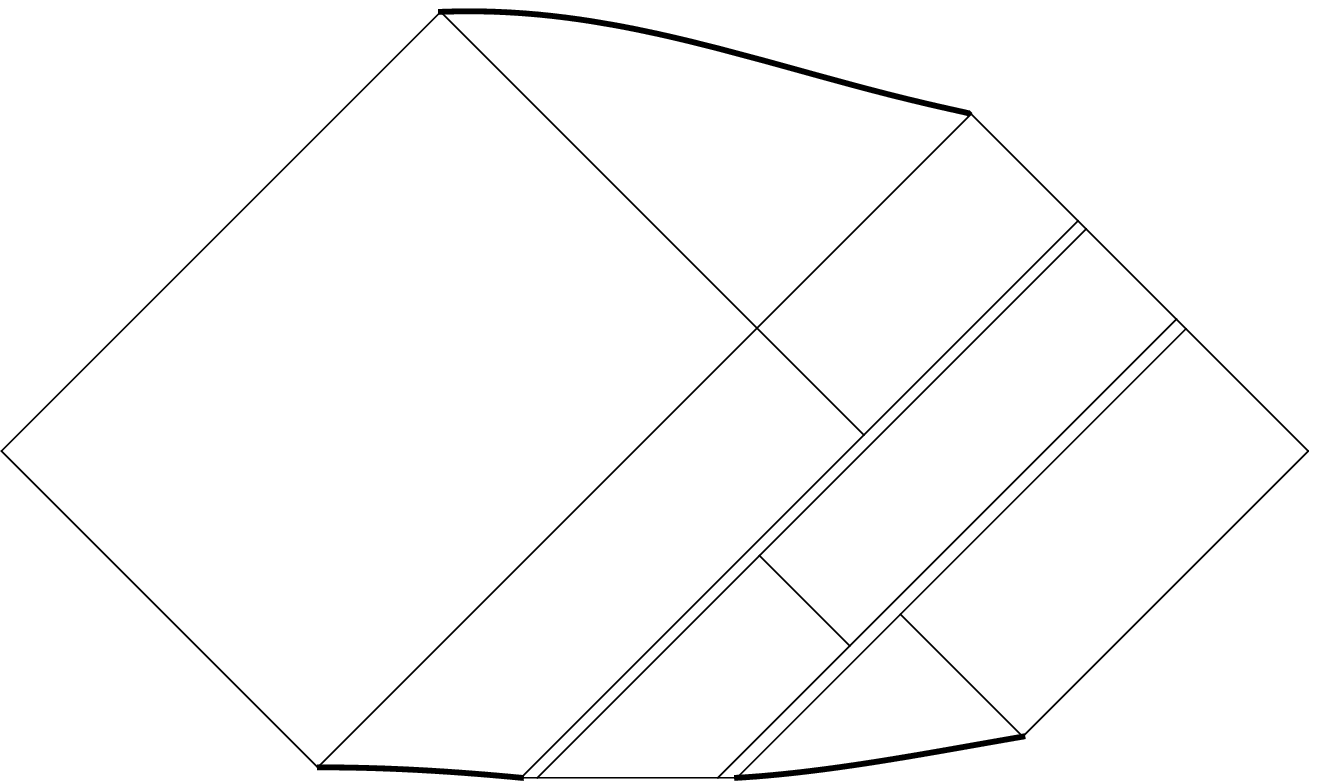}
\end{minipage}
\begin{minipage}{0.47\linewidth}
\includegraphics[width=3in]{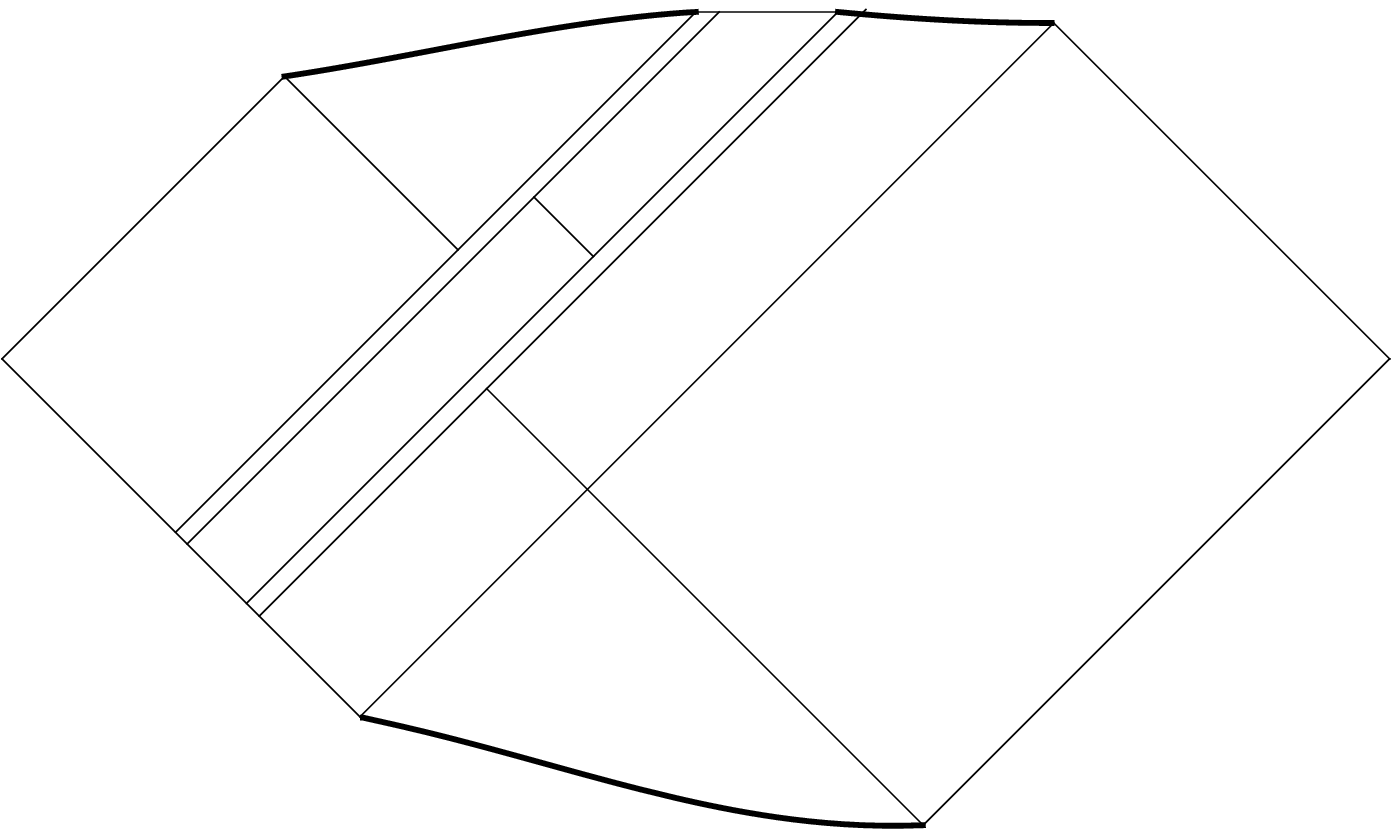}
\end{minipage}
\caption{\textbf{Two Shells with Equal but Opposite Total Energy:} These four
figures cover the four possible configurations of two concentric shells with
equal but opposite total energy, drawn in the $u_2$ and $v_2$ coordinates.
The first two figures represent concentric shells, where the inner shell has
positive energy density.  The last two figures are Penrose diagrams of
concentric shells where the inner shell has negative energy density, and the
overall shift allows for travel between the exterior Schwarzschild regions.}
\label{EqualShells}
\end{center}
\end{figure}

It is particularly interesting to consider concentric shells with equal but
opposite total energy.  Looking at Figure~\ref{EqualShells}, it is obvious
that there is a net shift in the horizons, even though the Region~1 and
Region~3 Schwarzschild regions have the same mass.  This net shift arises
because each shift is dependent on the energy density, not the total energy,
as can also be seen from the dependence of the stress-energy tensor on the
radius of the shell in~\eqref{Ttt}.  Since the inner shell will always be
smaller, and hence more dense, than the outer shell, there will always be a
net shift when the shells have equal but opposite total energy.
	
\section{Discussion}
\label{Discussion}

The primary purpose of this work has been to outline a method for drawing
accurate Penrose diagrams for Schwarzschild-like spacetimes containing
spherical shells of massless matter.  This method is straightforward: solve
the boundary conditions given in~\Dray, then construct particular diagrams
numerically.  This construction was originally motivated by the observation
that the diagrams given in~\Dray\ cannot in fact be accurate, as discussed in
Section~\ref{DtH}.  However, the accurate diagrams constructed in
Section~\ref{Shifted} do not in fact contain any new physical insight; one
could argue that the diagrams in~\Dray\ are indeed correct, even though some
artistic liberty was taken.

The situation is different when negative energy shells are considered.  First
of all, as discussed in Section~\ref{Negative}, even a single negative energy
shell leads to traversable wormholes; the shift goes ``the other way''.  Of
even greater interest, we showed in Section~\ref{Successive} that nested
shells of equal but opposite total energy still lead to traversable wormholes
--- provided the negative energy shell is inside the positive energy shell.
Thus, should ``nested'' positive/negative-energy particle pairs exist, they
would generate traversable wormholes!

How big would such a wormhole be?  As a characteristic scale, we determine the
``width'' between the transverse horizons in Regions~1 and~3.  As a measure of
this width, we compute the difference $\Delta t$ in Killing time between the
null extensions of those horizons into Region~2, measured along a null line
parallel to the shells.  For definiteness, we consider the situation shown in
the lower left diagram in~Figure~\ref{EqualShells}.  In null Kruskal-Szekeres
coordinates,
\begin{equation}
t=2m_2\log \left( -\frac{v_2}{u_2} \right)
\label{tKS}
\end{equation}
Inserting $v_1=0$ (respectively, $v_3=0$) on the transverse horizon in
Region~1 (respectively, Region~3) into~\eqref{UnequalV}, and then
into~\eqref{tKS}, using~\eqref{u2u1}, and subtracting, almost everything
cancels (since we are assuming $u$ is constant), and we are left with
\begin{equation}
\Delta t = 2m_2 \log \left( \frac{\alpha_1}{\alpha_2} \right )
\label{Dt}
\end{equation}
Thus, the ``width'' of the wormhole depends on the ``distance'' between the
shells, as given by the ratio of the $\alpha_i$, which can be chosen
arbitrarily.

Remarkably,~\eqref{Dt} depends on the background mass $M=m_1=m_3$ only through
the Schwarzschild mass of Region~2, which satisfies $m_2=M-m$, where $m$ is
the (magnitude of the) total energy in each shell.  If we assume that $m<<M$,
then $\Delta t$ scales linearly with~$M$.  Even if we choose the $\alpha_i$ to
be nearly the same, the effect might, in principle, be large enough to
measure.  For example, if $\frac{\alpha_1}{\alpha_2}=1.01$, $m$ is the
electron mass, and $M$ the mass of the sun, then $\Delta
t\approx29~\hbox{m}\approx10^{-7}~\hbox{s}$.

In summary, we have succeeded in constructing ``jigsaw puzzle pieces'' which
can be combined to analyze more complex arrangements of shells, such as the
successive shells considered in Section~\ref{Successive}.  Although new pieces
do need to be constructed to analyze scenarios with additional shells, this
method is still a useful tool for analyzing the effect of those shells.  In
future work, we hope to adapt these methods so as to be able to analyze
intersecting shells, as were also considered by~\Dray.

\section*{Acknowledgment}

This paper is an extension of the paper submitted by JSH in partial fulfillment
of the degree requirements for his M.S.\ in Physics at Oregon State
University~\cite{Hazboun}.

\end{document}